# Doping-Dependent Nonlinear Meissner Effect and Spontaneous Currents in High-$T_c$ Superconductors

Sheng-Chiang Lee, Mathew Sullivan, Gregory R. Ruchti,[a] and Steven M. Anlage
Center for Superconductivity Research, Physics Department, University of Maryland, College Park, Maryland 20742-4111

Benjamin S. Palmer
Laboratory for Physical Sciences, College Park, Maryland 20740

B. Maiorov[b] and E. Osquiguil
Centro Atómico Bariloche and Instituto Balseiro, Comisión Nacional de Energía Atómica, 8400 S. C. de Bariloche, Río Negro, Argentina

We measure the local harmonic generation from superconducting thin films at microwave frequencies to investigate the intrinsic nonlinear Meissner effect near $T_c$ in zero magnetic field. Both second and third harmonic generation are measured to identify time-reversal symmetry breaking (TRSB) and time-reversal symmetric (TRS) nonlinearities. We perform a systematic doping-dependent study of the nonlinear response and find that the TRS characteristic nonlinearity current density scale follows the doping dependence of the de-pairing critical current density. We also extract a spontaneous TRSB characteristic current density scale that onsets at $T_c$, grows with decreasing temperature, and systematically decreases in magnitude (at fixed $T/T_c$) with under-doping. The origin of this current scale could be Josephson circulating currents or the spontaneous magnetization associated with a TRSB order parameter.

---

[a] Current address: Physics and Astronomy Department, Johns Hopkins University, Baltimore, MD 21218
[b] Current address: Superconductivity Technology Center, Mail Stop K763, Los Alamos National Laboratory, Los Alamos, NM 87545

It is predicted that all superconductors have an intrinsic nonlinear Meissner effect (NLME).[1,2,3] Many experiments have been conducted to observe this effect in high-$T_c$ superconductors (HTSC).[4,5,6,7,8] Some of this work observed a linear-magnetic-field-dependent penetration depth at low temperature, in agreement with theory.[1,2] However, the quantitative, and some qualitative, details of the NLME signals in these experiments did not agree with the theory. This is most likely due to the presence of other, stronger, nonlinearities. In particular, most NLME experiments utilize sample geometries that induce large currents on poorly prepared and characterized edges and corners of the sample. This leads to extremely large edge currents, which can lead to vortex entry and nonlinear behavior that overwhelms the NLME.[9,10] Therefore, a measurement of the *local* nonlinear properties of superconducting samples without edge and corner effects involved is desired.

In addition, nonlinear mechanisms in superconductors are likely to be doping dependent. Experimental work[11,12] (including the one we present here) on doping-dependent 3$^{rd}$ order harmonics or intermodulation distortion in high-$T_c$ cuprates suggest the time-reversal symmetric nonlinearities are doping dependent. On the other hand, time-reversal symmetry breaking nonlinearities are also expected to be doping dependent in these materials. For example, Varma[13] proposed a doping dependent time-reversal symmetry breaking (TRSB) nonlinear mechanism, which involves micro currents flowing along the bonds in the $CuO_2$ planes in all under-doped high-$T_c$ superconductors for T < T* (the pseudo-gap temperature). An aspect of this proposal was tested by angular resolved photoemission spectroscopy,[14,15] but no consensus on the results has been achieved. Therefore, an independent technique for detecting nonlinear mechanisms in superconductors is desired, and we believe that our near-field nonlinear microwave microscope is a suitable alternative.

We employ a near-field microwave microscope to produce currents at high spatial frequencies, allowing us to detect small domains (e.g. TRSB domain sizes are predicted to be ~100 $\mu m$ )[13] that may have distinct Time-Reversal Symmetric (TRS) or TRSB nonlinearities. Measurements of the nonlinear properties of superconductors may give new insights into the basic physics of these materials.

The details of our microscope can be found elsewhere.[16,17,18] The basic idea is to send a microwave signal at frequency *f* to a local area of a superconducting film via the magnetic coupling between a loop probe and the sample surface, and to induce microwave currents in the sample far from the edges. The induced current distribution J(x,y) is concentrated on a lateral length scale of ~200 μm with maximum value ~ $10^4$ A/cm$^2$, dictated by the geometry of the probe and its height (12.5 μm) above the sample surface. Since the sample is nonlinear, it generates higher harmonic currents at 2*f*, 3*f*, etc. in response to the driving microwave signal. Note that the 2*f* and 3*f* signals are spatially distributed as $J^2$(x,y) and $J^3$(x,y) respectively, and are more sharply peaked than the driving current distribution. This means that the nonlinear response comes from even smaller area of the sample. The 2*f* signal implies the presence of TRSB nonlinearities, and 3*f* implies the presence of TRS nonlinearities, e.g. the NLME. These higher harmonic signals couple back to the microwave system, and are measured by a spectrum analyzer. The measurements are carried out inside a shielded environment consisting of two layers of mu-metal (high permeability at room temperature), and two layers of cryo-perm (high permeability at low temperatures). The sample is supported by a non-magnetic ultra-low-carbon steel base so that all measurements are carried out in nominally zero static magnetic field (< 1 $\mu G$ ).[17,18]

The harmonic generation is a very sensitive way to study superconducting nonlinearities. Harmonic measurements have been performed on superconducting films and crystals to study the microwave nonlinear behavior as a function of temperature and external magnetic field.[19,20,21,22,23] In addition, this method can be used to study the nonlinear response of specific superconducting structures, e.g. a superconducting bi-crystal grain boundary.[16,18,24] In the present experiment, we use this technique to study the local nonlinear response of homogenous superconductors as a function of doping.

Our samples are c-axis oriented YBa$_2$Cu$_3$O$_{7-\delta}$ (YBCO) thin films deposited on SrTiO$_3$ and NdGaO$_3$ substrates by pulsed laser deposition. The film thickness ranges from *t* ~ 100 nm to 200 nm. All samples were deposited as nearly optimally doped. After deposition, they were treated by re-annealing in different oxygen pressures at different temperatures to achieve the desired doping levels.[25,26] Their transition temperatures were determined by measuring the AC susceptibility (in the two-coil transmission geometry at 120 kHz with magnetic fields ~ 1 mT) after the re-annealing process. The hole concentration *x* is estimated using the measured $T_c$ and the universal formula,[27] $\frac{T_c}{T_c^{optimal}} = 1 - 82.6(x - 0.16)^2$ , where $T_c^{optimal}$ is taken as 93K, the highest $T_c$ we found from the literature. The summary of $T_c$ and transition width $\delta T$ of these samples along with the estimated hole concentration *x* is given in Table 1.

Table 1 Summary of $T_c$, the finite transition width $\delta T$ , and estimated hole concentration *x* of samples presented in this work.

| Sample | $T_c$ (K) | $\delta T$ (K) | Hole concentration x |
|--------|-----------|----------------|----------------------|
| MCS48  | ~ 46      | ~ 3            | 0.082                |
| MCS4   | ~ 54      | ~ 1.7          | 0.088                |
| MCS50  | ~ 75      | ~ 1.7          | 0.116                |
| MCS2   | ~ 84      | ~ 1.3          | 0.13                 |
| MCS3   | ~ 90      | ~ 0.7          | 0.16                 |

The nonlinear response of superconductors is temperature dependent and much can be learned about the microscopic properties through measurement of the temperature dependence of 2*f* and 3*f* nonlinear signals (referred to as $P_{2f}$ and $P_{3f}$). Shown in Fig. 1 is a typical $P_{2f}(T)$ and $P_{3f}(T)$ measured on an under-doped YBCO thin film. The AC susceptibility data shows a sharp transition temperature at ~ 75K. Both $P_{2f}(T)$ and $P_{3f}(T)$ show a peak near $T_c$. The peak in $P_{3f}(T)$ at $T \leq T_c$ is expected because the superfluid density is small and very sensitive to perturbations such as applied currents and magnetic fields.[3,28] The peak is well described by the NLME based on, e.g. the Ginzburg-Landau (GL) theory.[3,29] The inset to Fig. 1 shows a GL fit to the $P_{3f}(T)$ peak around $T_c$ in nearly optimally-doped YBCO with $T_c$ ~ 89.9K. This fitting uses the quadratically current-dependent superfluid density derived in Ref. [3] for type II superconductors,

$$\frac{n_s(T,J)}{n_s(T,0)} = \frac{\lambda^2(T,0)}{\lambda^2(T,J)} \cong 1 - \frac{1}{2}\left(\frac{J}{J_0(T)}\right)^2 \quad ,$$

where $J_0(T) = J_c(1-(T/T_c)^2)\sqrt{1-(T/T_c)^4}$ is the GL characteristic nonlinearity current scale, and $J_c$ is the depairing critical current density at T = 0. The superconductor is more sensitive to external perturbation (such as an applied current J) near $T_c$ because $J_0(T)$ goes to zero there. A finite transition temperature width $\Delta T$, cutoff screening length scale and finite $J_0(T_c)$ are used to smear out the divergence of the nonlinear response in the fitting,[30] and the parameters for the shown curve are $T_c$ = 89.9 K, $\Delta T$ = 0.45 K, and $J_c \sim 10^{11} A/m^2$.

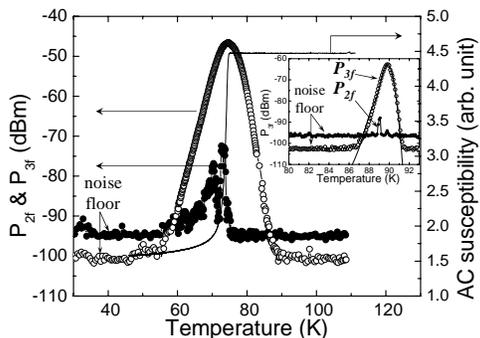

Fig. 1 A typical measurement of second and third harmonic response versus temperature in an under-doped YBCO thin film. The open circles are the $P_{3f}(T)$ data, solid circles the $P_{2f}(T)$, and solid line the AC susceptibility data. Both $P_{2f}$ and $P_{3f}$ show a peak near $T_c$, and only $P_{3f}$ extends to T > $T_c$. The inset is the harmonic measurement on an optimally doped YBCO fit to GL theory (solid line) with parameters $T_c$ = 89.9 K, finite transition width $\Delta T$ = 0.45 K, and $J_c$ ~ $10^{11}$ A/m².

However, we note that $P_{3f}(T)$ extends to temperatures substantially above $T_c$ in the under-doped samples, which is not expected from the ordinary NLME. It may be due to the enhanced fluctuations in under-doped cuprates, which leads to the appearance of residual $\sigma_2$ at high frequencies above $T_c$.[31,32,33] The present work is focused on the doping dependence of the nonlinear response near $T_c$, and future work will examine the temperature dependence.

While the $P_{3f}(T)$ data is semi-quantitatively understood (at least below $T_c$), the peak in $P_{2f}(T)$ is not expected from any theoretical proposals, to our knowledge. We performed a systematic study of $P_{2f}(T)$ and $P_{3f}(T)$ in under-doped YBCO thin films, and always observed this peak of $P_{2f}(T)$ near $T_c$, although it is smallest at optimal doping. Note that the experiment is performed inside four layers of magnetic shielding: two layers of mu-metal and two of cryo-perm. This shielding reduces residual magnetic fields and we find that the $P_{2f}(T)$ and $P_{3f}(T)$ peaks are highly reproducible under these conditions.[17] Unlike the peak in $P_{3f}(T)$, $P_{2f}(T)$ does not extend to T > $T_c$, and abruptly onsets above the noise floor at $T_c(x)$ in all samples.

Both the magnitude and the width of the $P_{2f}(T)$ and $P_{3f}(T)$ peaks near $T_c$ are systematically doping dependent. Fig. 2 is a summary of $P_{2f}(T)$ and $P_{3f}(T)$ of differently doped YBCO thin films, with the temperature scaled by $T_c(x)$. A general trend of enhanced $P_{2f}$ and $P_{3f}$ near $T_c$, and broader distribution of $P_{2f}(T/T_c)$ and $P_{3f}(T/T_c)$ in the more under-doped YBCO is observed.

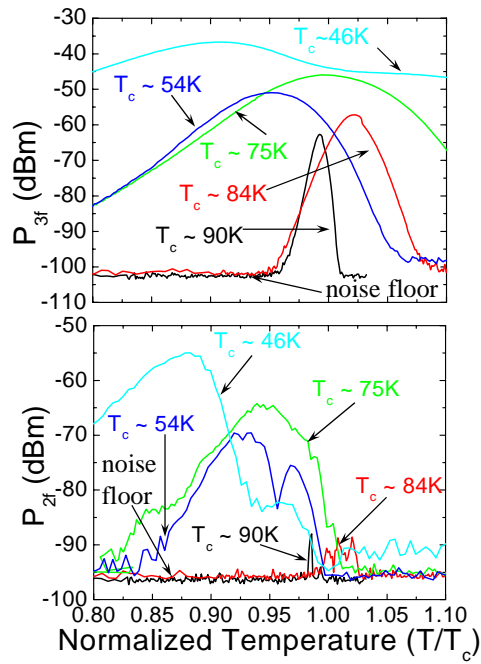

Fig. 2 Summary of $P_{2f}(T)$ and $P_{3f}(T)$ data measured from variously doped YBCO thin films. The temperature is normalized by the $T_c$ determined by AC susceptibility. The measured transition widths in ac susceptibility of these samples are shown in Table 1.

To quantitatively understand the doping dependence of nonlinearities from the harmonic data,

we need to turn our data into a detail-independent measure that directly reflects the nonlinear mechanisms. It is generally accepted that the current-dependent super-fluid density for the NLME can be generalized to describe the nonlinear behavior of other time-reversal symmetric (TRS) mechanisms as,[34,35]

$$\frac{n_s(T,J)}{n_s(T,0)} = \frac{\lambda^2(T,0)}{\lambda^2(T,J)} \cong 1 - \left(\frac{J}{J_{NL}(T)}\right)^2,$$

where $n_s$ is the super-fluid density, $\lambda$ is the magnetic penetration depth, and $J_{NL}$ is the nonlinear scaling current density, which depends on the nonlinear mechanisms (e.g. the de-pairing critical current density for the NLME). The expansion holds when $J/J_{NL}(T) \ll 1$. The scaling current density quantitatively determines the degree of nonlinearity of the associated mechanism. The smaller the value of $J_{NL}$, the more nonlinear the associated mechanism.

Since the electromagnetic response of a superconductor is primarily inductive in nature, it is reasonable to assume that the reactance of a superconductor dominates its nonlinear response. Following an algorithm described in detail elsewhere,[16,17,18,35] we calculate the current-dependent superconducting inductance $L = L_0 + L_2 I^2$, and the third harmonic response generated by $L_2$. The power, $P_{3f}$, is proportional to the square of the third harmonic voltage developed on this nonlinear inductor, and depends on the penetration depth, applied power ($P_f$), frequency, $J_{NL}(T)$, and the geometry of the driving current distribution $J(x,y)$. With this, we can extract the scaling current density from the $P_{3f}(T)$ data as,[16,17,18]

$$J_{NL}(T,x) \cong \sqrt{11.94\Gamma \times \frac{\omega\mu_0\lambda^2(T,x)}{4t^3\sqrt{2Z_0 P_{3f,measured}(T,x)}}},$$

where $\Gamma$ is the geometry- and probe-dependent figure of merit of the microscope, $\Gamma \sim 31$ A$^3$/m$^2$ at +12 dBm input microwave power, $\omega/2\pi \sim 6.5$ GHz is the driving frequency, $Z_0 = 50$ $\Omega$ is the characteristic impedance of coaxial transmission lines, $t$ is the film thickness, and the numerical factor accounts for calibrated system-specific details (amplification, cable attenuation, etc.). Since the penetration depth $\lambda$ is involved in the algorithm, to extract the doping-dependence of $J_{NL}$ from experimental harmonic data we must consider the doping dependent penetration depth ($\lambda(T,x)$ data is taken from the literature).[36,37,38,39]

Shown in Fig. 3 is the $J_{NL}$ extracted from the $P_{3f}(T)$ data shown in Fig. 2 at a temperature of 0.97$T_c$.[40] A trend of decreasing $J_{NL}$ in the more under-doped YBCO films is observed. If the dominant nonlinear mechanism which gives rise to the $P_{3f}$ signal near $T_c$ is the NLME, then the extracted $J_{NL}$ should be a measure of the de-pairing critical current density, $J_c$.

The de-pairing critical current density $J_c$ is related to the thermodynamic critical field $H_c$, and the condensation energy density $U$ of the superconductor as

$$J_c(T,x) \approx H_c(T,x)/\lambda(T,x) = \sqrt{2U(T,x)/\mu_0}\Big/\lambda(T,x).$$

Taking the doping dependent penetration depth into account, the observed doping dependent $J_{NL}$ implies that the condensation energy is also doping dependent. This statement is supported by the work of Luo et al.[41] who measured the doping dependent zero-temperature condensation energy density $U(0, x)$ of YBCO ceramics. We extract $J_c(T = 0)$ from their data and plot with our doping dependent $J_{NL}$ in Fig. 3 for comparison, where the doping trends are in good agreement.[42] This suggests that we have measured the doping dependence of the NLME nonlinear scaling current density in YBCO.

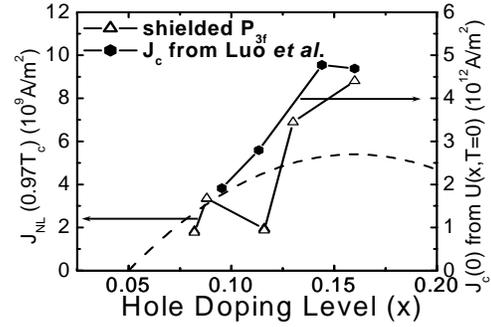

Fig. 3 Plot of $J_{NL}(T = 0.97T_c)$ extracted from $P_{3f}$ data in Fig. 2 vs. the doping level. The doping levels are estimated from $T_c$ using the universal formula mentioned in the text and are summarized in Table 1. The depairing critical current density at zero temperature $J_c(T = 0)$ extracted from the zero-temperature condensation energy by Luo et al.[41] is plotted for comparison. The dashed line schematically illustrates the superconducting dome in the phase diagram.

The observation of a $P_{2f}$ signal near $T_c$ implies the presence of TRSB nonlinearities and a spontaneously flowing current in or on the superconductor. To obtain a quantitative understanding of the measured $P_{2f}$ signals, we phenomenologically propose that the spontaneous current simply modifies the NLME in a way leading to broken time-reversal symmetry:

$$\frac{n_s(T,J)}{n_s(T,0)} = \frac{\lambda^2(T,0)}{\lambda^2(T,J)} \cong 1 - \left(\frac{J - J_{TRSB}(T)}{J_{NL}(T)}\right)^2,$$

$$\cong 1 + \frac{2J\,J_{TRSB}(T)}{J_{NL}^2(T)} - \left(\frac{J}{J_{NL}(T)}\right)^2,$$

where $J_{TRSB}$ is the phenomenological spontaneous current density. Rewriting the linear term as $J/J_{NL}'$, where $J_{NL}' \equiv J_{NL}^2/2J_{TRSB}$, we can follow a similar analysis algorithm[16,17,18,35] to extract $J_{NL}'$, hence $J_{TRSB}$. As a result, we find that $J_{TRSB}$ *can be extracted directly from the data* without considering the doping dependence of $\lambda$, or the precise value of $T_c$ as

$$J_{TRSB}(T) \cong \frac{2.8(A/m)}{t}\sqrt{\frac{P_{2f,measured}(T)}{P_{3f,measured}(T)}}.$$

The extracted $J_{TRSB}(T)$ of YBCO films at different doping levels are shown in Fig. 4. Note that the data can only be presented in a temperature range where both $P_{2f}$ and $P_{3f}$ signals are above the noise floor, i.e. $0.97 < T/T_c < 0.99$. There are two trends observed in this figure. First is that $J_{TRSB}(T)$ of all under-doped YBCO films appears to onset at $T_c$, which can also be concluded from the $P_{2f}(T)$ data in Fig. 2. From the temperature dependent $J_{TRSB}$ shown in Fig. 4, it is likely that there is no $J_{TRSB}$ above $T_c$ in striking contrast with the third harmonic (TRS) response. Secondly, the magnitude of $J_{TRSB}$ at a fixed $T/T_c$ generally decreases in the more under-doped films. To understand these behaviors, we must identify the origin of the nonlinear mechanism.

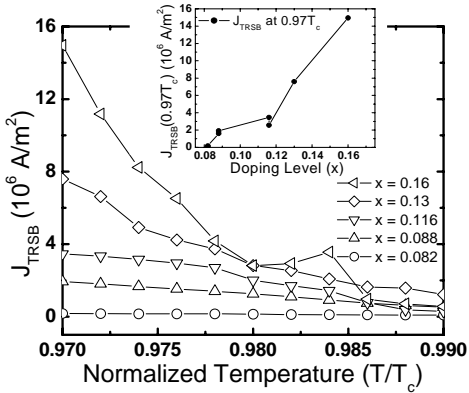

Fig. 4 The main figure is $J_{TRSB}(T/T_c)$ extracted from harmonic measurements of variously doped YBCO films. The temperature is normalized by $T_c$ measured by ac susceptibility. The inset is a plot of $J_{TRSB}(T = 0.97 T_c)$ vs. the doping level to show the doping dependence of $J_{TRSB}$.

One possible mechanism for TRSB nonlinearities is Josephson vortices in a superconducting weak-link network. It is known that high-$T_c$ cuprates, and particularly under-doped YBCO,[26] are often granular in nature. Although it is proposed that the superconducting order parameter onsets at much higher temperatures (pseudo-gap temperature) in under-doped cuprates, the long-range phase coherence is not established until the temperature reaches $T_c < T^*$.[43] This is a necessary condition for the existence of Josephson vortices in the weak-link network since long-range superconducting currents are required to circulate among the network.

This argument suggests that the magnitude of $J_{TRSB}$ ($<< J_{NL}$ at the same $T/T_c$) is on the order of the Josephson critical current density ($10^7$ A/m$^2$ at $T/T_c = 0.97$) of the weak-link network. On the other hand, the doping dependence of $J_{TRSB}$ implies that the Josephson critical current density is doping dependent. This statement is supported by the work of Sydow et al.,[44] who measured the Josephson critical current of 23° YBCO bi-crystal grain boundaries at different doping levels. They found that the critical current drops by a factor of ~ 100 from an optimally doped junction to an under-doped junction of $T_c$ ~ 50K. We expect that the low-angle grain boundaries, which may be present in our films, demonstrate similar doping dependence.

However, power-dependent measurements of $P_{2f}$ near $T_c$ are not entirely consistent with a Josephson vortex mechanism for $P_{2f}$. We observe a monotonic and uniform power-law dependence of $P_{2f}(P_f) \sim P_f^{1.8}$, as shown in Fig. 5. This is not expected from a Josephson nonlinearity, which should reveal a non-monotonic power dependence of $P_{2f}(P_f)$, followed by saturation, as we have observed with the same microscope in an isolated YBCO bi-crystal grain boundary,[16] and which is also expected from simulations.[45] The uniformity of the $P_{2f}(P_f)$ response suggests that a global nonlinear mechanism may be responsible for the observed signal.

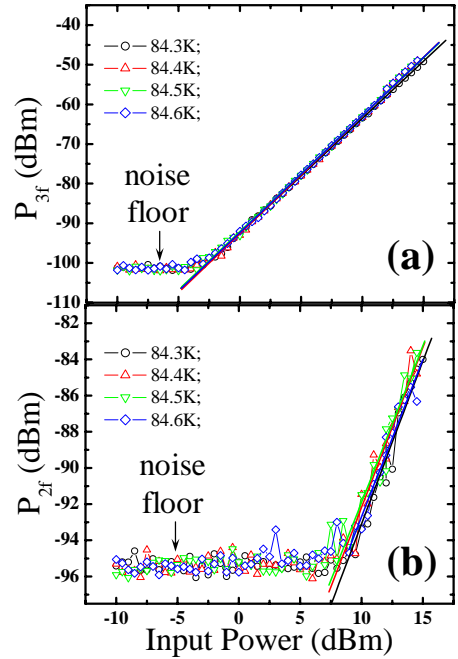

Fig. 5 The power-dependent (a) $P_{3f}$ and (b) $P_{2f}$ measured for the YBCO thin film with $x = 0.13$ and fit at temperatures near $T_c$ ~ 84 K. All axes are on a logarithmic scale. The solid lines are power-law fits with slopes of close to 3 and 2 for $P_{3f}$ and $P_{2f}$, respectively, which is consistent with assumptions used in the algorithms for extracting $J_{NL}$ and $J_{TRSB}$ discussed in the text.

The power-dependence of $P_{2f}$ is consistent with another possible mechanism for $J_{TRSB}$, namely the presence of an exotic TRSB order parameter associated with a temperature-dependent spontaneous magnetization.[13,46] This order parameter is expected to have a spontaneous magnetic field and current associated with it when translational symmetry is also broken.[47] Several possibilities for microscopic TRSB surface states in YBCO include $s + id_{x^2-y^2}$ and $d_{x^2-y^2} + id_{xy}$-wave pairing states, which may break into domains of degenerate states.[46] We note that the behavior of $J_{TRSB}(T)$ in Fig. 4 resembles that of a TRSB order parameter, such as that seen as an internal magnetic field in $Sr_2RuO_4$[48] by muon spin relaxation ($\mu$-SR), for example. This suggests the possibility of a TRSB order parameter onset at $T_c$ in under-doped cuprates. Also, several theorists have pointed out that an observation of fractional vortices with flux other than an integer or half-integer flux quanta would be an indication of broken time-reversal symmetry.[46,49] The experimental work of Tafuri and Kirtley[50] observed fractional vortices in c-axis YBCO films by scanning SQUID microscopy, consistent with this picture. They reported the enhancement of the local magnetization associated with these vortices as the temperature is decreased, consistent with our observation of larger $J_{TRSB}$ at lower temperatures.

Note that the smallest measured magnetic field associated with $J_{TRSB}$ can be estimated as $B \cong \mu_0 J_{TRSB} t \leq 0.1 mG$, where $t \sim 1000 \text{Å} < \lambda$ is the film thickness and $J_{TRSB} \sim 10^5$ A/m$^2$ is the smallest $J_{TRSB}$ we measured. Note that the sensitivity limit to the local magnetic field claimed by $\mu$-SR is about 100 mG[48]. This suggests that our microscope has excellent sensitivity to TRSB mechanisms in surface states and thin films. We also note that our microscope generates nonlinearity-induced current distributions on the length scale of at least one set of proposed TRSB domains.[13]

In conclusion, we have demonstrated direct measurements of the NLME near $T_c$ of YBCO films at different doping levels, and found the de-pairing critical current density to decrease with the doping level below optimal. This is the first measurement of the doping dependence of the NLME, to our knowledge. We also observe a temperature and doping-dependent TRSB nonlinearity from $P_{2f}$ and $P_{3f}$ measurements, which may be due to the presence of Josephson vortices in a weak-link network, or to the presence of a TRSB order parameter associated with domains of spontaneous magnetization. We phenomenologically introduce a spontaneous current $J_{TRSB}$ to quantify the strength and doping dependence of this nonlinearity. Our results are free of edge effects, and have shown the unique ability of our technique to study weak local nonlinearities of superconductors.

We thank Juergen Halbritter for useful discussions. This work is supported by NSF/GOALI DMR-0201261, and the Microwave Microscope Shared Experimental Facility of the NSF/Maryland MRSEC DMR-0080008.